\documentclass[aps,prb, twocolumn,floatfix,showpacs]{revtex4-1}
\usepackage{graphicx}
\usepackage{amsfonts,amsmath,amssymb}
\usepackage{amsthm}
\usepackage{dsfont,bm}
\usepackage{color,soul}
\usepackage{xcolor, soul}
\sethlcolor{yellow}
\usepackage{amsbsy}
\usepackage[colorlinks=true,linkcolor=blue,pagecolor=blue,filecolor=blue,menucolor=blue,urlcolor=blue,citecolor=blue,anchorcolor=blue]{hyperref}%

\usepackage{mathbbol}

\usepackage{sidecap}

\begin{document}

\title{Anisotropic angle-dependent Andreev reflection at the ferromagnet/superconductor junction on the surface of topological insulators}


\author{Morteza Salehi}
\affiliation{Department of Physics, Bu-Ali Sina University, Hamadan 65178, Iran}

\begin{abstract}
We theoretically demonstrate that a ferromagnetic/superconductor junction on the surface of three-dimensional topological insulators (3D TIs) has an anisotropic angle-dependent Andreev reflection when the in-plane magnetization has a component perpendicular to the junction. In the presence of in-plane magnetization, the Dirac cone's location adjusts in the $k$-space, whereas its out-of-plane component induces a gap.
This movement leads to the anisotropic angle-dependent Andreev reflection and creates transverse conductance flows parallel to the interface. Also, an indirect gap induces in the junction, which removes the transport signatures of Majorana bound states. Because of the full spin-momentum locking of Dirac fermions on the surface of 3DTIs,  a torque that called \textit{Andreev Transfer Torque} (ATT) imposes on the junction. Moreover, we propose a setup to detect them experimentally.

\end{abstract}


\maketitle

\section{Introduction}
Since spin-transfer torque (STT) lives at the heart of the data storage industry, it attracts great interest\cite{Chappert20099,Parkin 2008,Pinarbasi,Mamura2022JMMM,Tsymbal2009Book,Cai2021Sci,Yuasa2018}. The mutual interactions between the spin of charge carriers and magnetic orders cause the STT. This torque originates on the transfer of spin angular momentum
of spin current to the magnetization or vice versa\cite{Bazaliy1998PRB,Tsoi19998PRL,Stiles2002PRB}. This effect works in electronic devices such as oscillator circuits or magnetic random access memory\cite{Zutic2011,Cui2022Spintronics}. The heat dissipation due to electric resistance is one of the most critical issues in spintronics. The dissipationless current in superconductors in combination with ferromagnets proposes new types of devices to manipulate spin and charge currents.  \cite{Linder2015NatPhys,Shomali,Moen2018PRB,Bobkova,Haltermann}.

The interface of superconductors can reflect the incoming electron from the non-superconducting side as a backscattered hole while a Cooper pair enters the superconductor\cite{Andreev1964JETP}. This process, known as Andreev reflection, is dominant at voltages below the superconducting gap. During the Andreev reflection, the dissipative current converts to the dissipationless current\cite{BTK}. 
The reflected hole will be created in the conduction band when its energy excitation is less than chemical potential. This hole moves back alongside its incident electron in real space, known as retro-reflection. For an incident electron with energy bigger than the chemical potential, the corresponding hole locates in the valence band during the Andreev process. This hole, similar to an optical ray in front of a mirror, moves back specularly\cite{Beenakker2006PRL,Zhang2008PRL,Schelter2012PRL}. 

On the other hand, topological insulators (TIs) are a class of materials with non-trivial properties, first proposed theoretically and then confirmed experimentally\cite{Kane2005PRL,Kane2005PRL-2,Fu2007PRB,Bernevig2006Sci,Fu2007PRL,Teo2008PRB,Hsieh2008Nau,Hsieh2009Nat,Hsieh2009Sci,Zhang2009NP,Kuroda2010PRL,Hazzan2010RMP,Jackiw1976PRD}. Due to the bulk-edge correspondence, they have gapless states on their edges or surfaces\cite{Jackiw1976PRD}. In contrast to the Graphene\cite{Novoselov2004Sci,Novoselov2005Nature}, these states are fully spin-orbit coupled and protected against local perturbations\cite{Hazzan2010RMP}. A Dirac-like Hamiltonian governs on the carriers in the low excitation approximation \cite{Dirac1928PRSL}.
In the presence of spin-orbit interaction, the spin operators could not be well-defined for metallic materials\cite{Rashba2003PRB,Soori,Soori2}, whereas chirality can be a well-defined operator for TIs because of strong spin-momentum locking\cite{Beiranvand2021JOP}.
Also, magnetization and superconductivity can be induced in these states\cite{Tikhonov2016PRL,Cheklesky2012NP,Cheklesky2014NP,Qi2009Sci} to make them an exciting platform for exploring new phenomena\cite{Yokoyama2010PRB,Yokoyama2009PRL,Linder2010PRL,Linder2010PRB,McIver2011NN,Salehi2011}. 
In our previous work, unlike early attempts to explore the physics of STT on TIs, we focus on low energy excitation regime to reveal the Dirac physics. We show a current transfer torque imposed on the ferromagnet/ normal (F/N) junction of three-dimensional topological insulators (3D TIs) which can be detected via a Hall voltage\cite{Beiranvand2021JOP}. 
So far, the STT of superconducting-based devices is considered without focus on the role of Andreev reflection. 

\begin{figure}
\includegraphics*[scale=0.4]{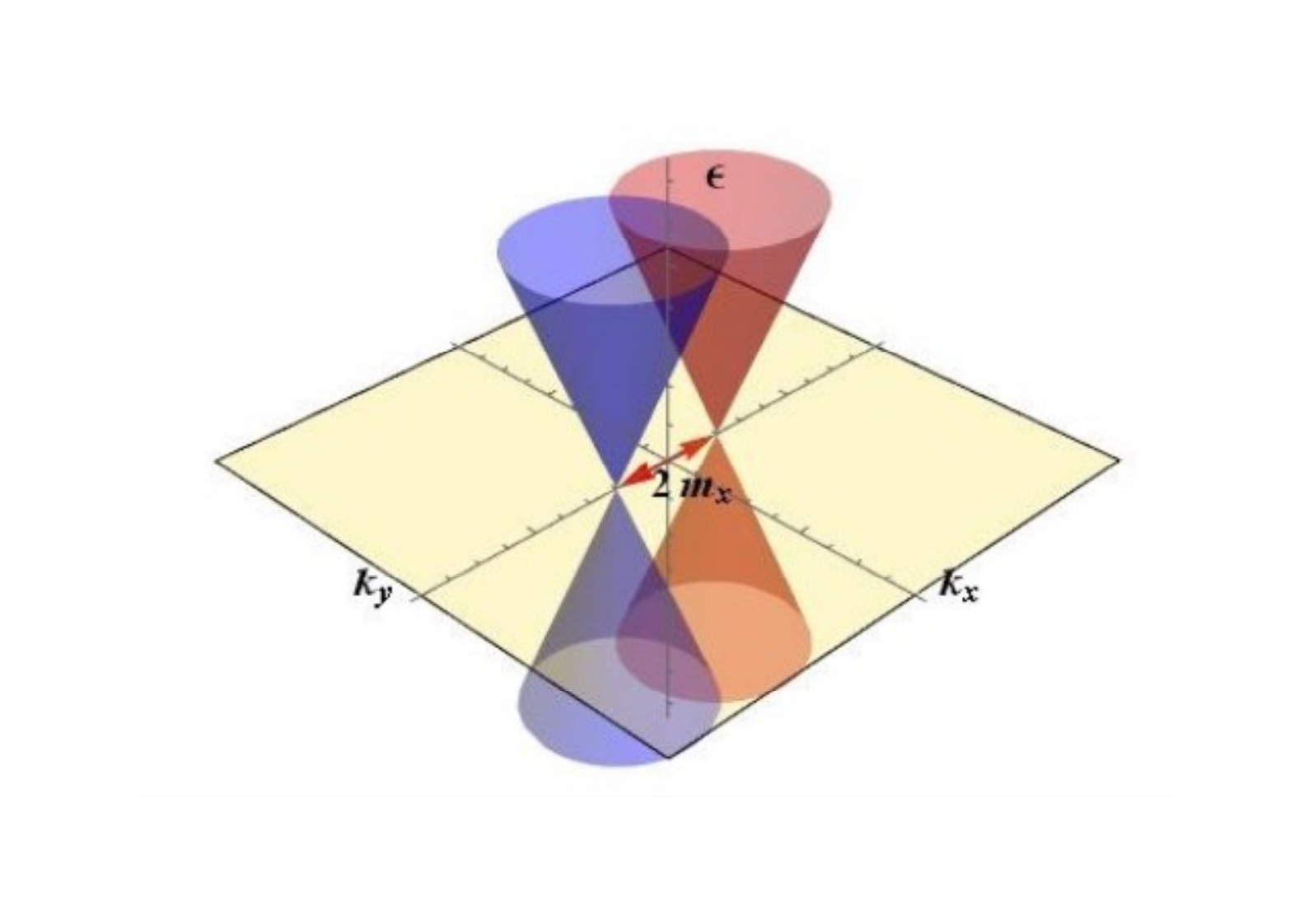}
\caption{(color online) The dispersion relation of electron-like (blue cone) and hole-like (red cone) quasi-particles in the $k$-space, respectively. In the presence of in-plane magnetization, two cones separate in the $k$-space with $2 \sqrt{(m_x^2+m_y^2)}$. Here, we set $m_y=\mu=0$. The upper part of blue cone is empty for electron-like excitations whereas the upper part of red cone is filled for hole-like excitations.}
\label{Fig.Dispersion}
\end{figure} 
To upgrade our theory to explore the importance of Andreev reflection, we consider a ferromagnet/superconductor (F/S) junction on the surface of 3D TIs, where the proximity effect induces magnetization and superconductivity. We assume superconductivity has an s-wave character and the propagation of Dirac fermions occurs in the ballistic regime. To explore it theoretically, we use the Bogoliubov-deGennes (BdG) equation\cite{deGennes1999Book},

\begin{equation}
H_{BdG}=\left(
\begin{array}{cc}
H_D(k)-\mu & \Delta \\
\Delta^{*}&  \mu-\mathcal{T}H_D(k)\mathcal{T}^{-1}\\
\end{array}\right).
\label{Eq.HBdG}
\end{equation}

Here, $H_D(k)$ is the effective Hamiltonian that governs on the Dirac fermions of the surface of 3D TI in the presence of magnetization. It can be written as,

\begin{equation}
H_D(\textbf{k})=\hbar v_F (\boldsymbol{\sigma}\times \textbf{k}).\hat{e}_z-( \textbf{m}. \boldsymbol{\sigma})
\label{Eq.Hamiltoina},
\end{equation}

where $\boldsymbol{\sigma}$ and $\textbf{k}$ are Pauli spin vector and  wave vector, respectively. Also, $\textbf{m}=\textbf{m}_0 \Theta(-x)$ is the effective magnetization coupled to the spin degrees of freedom and $v_F$ is the Fermi velocity. To avoid complexity, we set $\hbar v_F=1$ in the remainder of the paper. Moreover $\mathcal{T}=i\sigma_y K$ is the time-reversal operator and $K$ is the complex one. Moreover, $\Delta=\Theta(x)\Delta_0 \sigma_0$ is the complex superconducting order parameter and $\sigma_0$ is a $2\times 2$ unit matrix in the spin space. Here, $\Theta(x)$ is the Heaviside step function. The $\mu$ stands for the chemical potential that can be tuned by external gate. In the absence of superconductivity, the excitation energy of electron-like and hole-like quasi-particles according to Eq.(\ref{Eq.HBdG}) are, 

\begin{eqnarray}
\epsilon_e=\pm \sqrt{(k_x+m_y)^2+(k_y-m_x)^2+m_z^2}-\mu,
\label{Eq.Dispersion1}  \\
\epsilon_h=\pm \sqrt{(k_x-m_y)^2+(k_y+m_x)^2+m_z^2}+\mu
\label{Eq.Dispersion2}
\end{eqnarray}.

Here, the $\{x,y,z\}$ indices belong to the space components of magnetization and wave vector. The Eq.(\ref{Eq.Dispersion1}) and Eq.(\ref{Eq.Dispersion2}) show two cones of electron-like and hole-like quasi-particles in the $k$-space, respectively. In the absence of magnetization, these cones initially locate at the center of the Brillouin zone. As shown in Fig.(\ref{Fig.Dispersion}), in-plane magnetization, $\{m_x \neq 0, m_y\neq 0, m_z=0\}$,  tunes the Dirac cone's location and separate them from each other with $2\sqrt{m_x^2+m_y^2}$\cite{Yokoyama2009PRL}. Also, out of plane magnetization induces a direct gap for both cones\cite{Linder2010PRB}.
The blue cone demonstrates the electron-like dispersion, whereas the red cone indicates on the hole-like one. An electron fills an empty state above the chemical potential at zero temperature. This electron moves in the real space along its group velocity, $\langle V_i \rangle_g=\langle \partial \epsilon / \partial k_i\rangle$. During the Andreev reflection, a hole is created on the hole cone's empty part. Since the parallel component of the wave vector and energy are conserved during the scattering processes, the cone's separation leads to angle-dependent Andreev reflection. It means the probability of Andreev reflection depends on the propagation direction of the incoming particle. 
This effect imposes a torque on the junction called  \textit{Andreev transfer torque} (ATT). Only the $z$-component of ATT is non-zero. Because of strong spin-orbit interaction on Dirac fermions of 3D TIs, a transverse current flows parallel to the interface of the F/S junction. This current can be detected via a four-terminal setup as its experimental signature.  

This paper organized as follows. In Sec.\ref{Sec.Theory}, we
demonstrate the physics of angle-dependent Andreev reflection due to the separation of Dirac cone's location in the $k$-space. The transport properties and the
continuity equation of spin density wave are calculated,
too. In the steady-state approximation, we show
that only the $z$-component of ATT is non zero. In Sec.\ref{Sec.Re1}, we illustrate the creation of an indirect gap in the transport probabilities. Since the Majorana-bound states locate at the zero energy on the interface of the F/S junction, this gap removes their signatures on the transport properties. Also, we show that transport probabilities are dependent on the propagation direction of incoming particles in the real space. In Sec.\ref{Sec.Re2}, we show a transverse current flows parallel to the interface. This effect is related to the direction of in-plane magnetization. In Sec.\ref{Sec.Re3}, the $z$-component of ATT is calculated. Finally, the conclusion is given in Sec.IV

\begin{figure}
\includegraphics[scale=0.26]{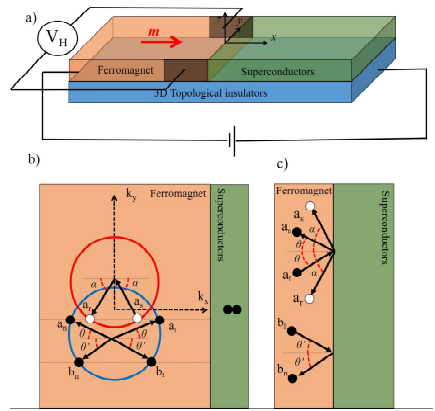}
\caption{(a) The schematic illustration four-terminal 3D TI-based ferromagnet/superconductor junction that can be used in experiment to detect ATT and transverse conductance. (b) The Andreev reflection processes are shown in $k$-space configuration in the presence of in-plane magnetization. (c) The corresponding real space processes of Andreev reflection, both for specular and retro reflections.}
\label{Fig.1}
\end{figure}

\section{Theory and formalism}
\label{Sec.Theory}

As illustrated in part (a) of Fig(\ref{Fig.1}), we consider a 3D TI-based four terminal F/S junction. The in-plane magnetization induces by means of proximity in the $x < 0$ part of the junction.
In the ballistic regime, the energy and parallel component of wave vector with respect to the junction are conserved during the scattering processes.  As shown in Fig.(\ref{Fig.Dispersion}), the $x$-component of magnetization moves the Dirac point toward $k_y$-direction whereas its $y$-component moves the Dirac point in $k_x$-direction of the $k$-space. Since the $x$-component causes the relative motion of reflected holes in Andreev processes, we assume the magnetization applied perpendicular to the junction,$\{m_x=-m_0, m_y=0, m_z=0\}$, where $m_0$ is the magnitude of magnetization.
 
A top view of TI-based F/S junction in the $k$-space is shown in part (b) of Fig.(\ref{Fig.1}). The blue circle belongs to the empty states of electron-like cone with energy $\epsilon$ higher than chemical potential, $\mu$. The radius of the electron-like circle is $\mu+\epsilon$. The black dot, labeled by ($a_I$), stands for an incoming fermion. Also, the black arrow centered at the origin of the blue circle determines the propagation direction of incoming fermion in the real space via its group velocity.  Since energy is conserved, the radius of empty states at hole-like circle (the red one) is equal to $|\mu-\epsilon|$. Moreover, the $k_y$ is conserved during the scattering processes. Using these conditions, the incoming fermion with energy $\epsilon \leq \Delta_0 $ encounters two scenarios:

\begin{itemize}
 \item In the Andreev zone where the electron-like and hole-like circles overlap, the incoming fermion can find an empty state with the probability of $R_A$ on the hole-like circle. If the hole-like circle locates at the valence band, the hole reflects specularly. This effect is shown by the white dot labeled by ($a_s$) in part (b) of Fig.(\ref{Fig.1}). On the other hand, the reflected hole in the conduction band belong to the retro-reflection type that is illustrated by the white dot labeled by ($a_r$) in part (b) of Fig.(\ref{Fig.1}). Also, there is a probability of $R_N$ for the incoming fermion to be reflected usually, which called normal reflection. It is shown by ($a_n$) in part (b) of Fig.(\ref{Fig.1}). The probability conservation ensures $R_N+R_A=1$. The black arrows centered at the origin of electron-like and hole-like circles determine the propagation directions of incoming and back-scattered states in the real space. This effect is shown schematically in the real space in the upper part of part (c) in Fig.(\ref{Fig.1}).

\item In the reflection zone where electron-like and hole-like circles do not overlap, there is no state on the hole-like circle for the incoming fermion.
It means the perfect normal reflection occurs, $R_N=1$. This effect is shown by label of ($b_I$) for the incoming fermion and ($b_n$) for the reflected one in part (b) of Fig.(1). Also,  this effect is depicted schematically in the real space in the lower part of part (c) in Fig.(\ref{Fig.1}).
\end{itemize}

Incoming fermions with energy $\epsilon > \Delta_0$ have another possibilities. They can transport across the junction to the states that exist above the superconducting gap. 
 
These different scenarios show that the junction is sensitive to the propagation direction of incoming particles due to the moving of the Dirac cones in the k-space. The incoming fermions located at the upper part of the electron-like circle can be reflected with the probability of $R_A$ in the Andreev reflection processes. In contrast, the incoming fermions located at the lower part have to be reflected normally. This process causes an anisotropic angle-dependent Andreev reflection.  Since incoming fermion and its reflected hole move in different directions, the net current they carry differs. So, it creates a transverse current that flows parallel to the interface and can be detected via a four-terminal setup. Due to the strong spin-momentum locking on the surface of 3D TIs, any change in the propagation direction of carriers causes a change in the spin configuration. This effect manifests itself through a torque that imposes on the junction.

We define the basis of wave functions of Eq.(\ref{Eq.HBdG}), as 
$\psi=(\phi_\uparrow, \phi_\downarrow, \phi^*_\downarrow, -\phi^*_\uparrow)^T$.  The $\uparrow (\downarrow)$ stands for up (down) spin direction.
The corresponding wave functions of electron-like excitations, Eq.(\ref{Eq.Dispersion1}), are derived such as,
\begin{equation}
\psi^\pm_e(r)=\left(\begin{array}{c}
i \\
\pm e^{\pm i \theta_e}\\
0\\
0\\
\end{array}\right)e^{\pm i \textbf{k}_e.\textbf{r}},
\label{Eq.Psie}
\end{equation}
where the $\pm$ sign refers to the right or left propagation direction and $\textbf{k}_e=(k^e_x,k_y)$ is the two dimensional wave vector of carrier. The eigenvalues of hole-like excitations is determined by Eq.(\ref{Eq.Dispersion2}) and their wave functions are,
\begin{equation}
\psi^\pm_h(r)=\left(\begin{array}{c}
0 \\
0\\
\pm \mathcal{S} e^{\pm i \alpha}\\
i\\
\end{array}\right)e^{\pm i \textbf{k}_h.\textbf{r}}.
\label{Eq.psih}
\end{equation}

 Here, $\mathcal{S}=sign(\epsilon-\mu)$ determines the location of hole-like excitation in the valence or conduction band. The $\alpha$ refers to the propagation direction.
 The $\textbf{k}_h=(k^h_x,k_y)$ is its wave vector. Since $k_y$ is conserved during the scattering process, we obtain from Eq.(\ref{Eq.Dispersion2}),
 
 \begin{equation}
 k^h_{x}=\pm \sqrt{(\epsilon-\mu)^2-(k_y+m_x)^2}+m_y
 \label{Eq.Kh}.
 \end{equation}
 This relation indicates that the reflected holes must satisfy $  |\epsilon-\mu|\geq |k_y+m_x| $ condition to find an stable state in hole-like cone.
 
 In the presence of superconductivity, $\Delta_0 \neq 0$, the wave functions of electron-like and hole-like excitations can be derived in a similar way. The group velocity operator is
\begin{equation}
\hat{V}= -\eta_0\sigma_y \hat{i}+\eta_0\sigma_x \hat{j},
\end{equation}
where $\eta_0$ is a $2 \times 2$ unit matrix in the Nambu space. One can derive the propagation direction of electron-like excitation in the F region such as,
\begin{equation}
\begin{array}{cc}
\langle \hat{V}_{e,x}\rangle=\cos \theta, & \langle \hat{V}_{e,y}\rangle=\sin\theta.
\end{array}
\end{equation}
The $\{\theta, \alpha \}$ are depicted in Fig.(\ref{Fig.1}). Also, the probability current density is $\textbf{J}=\langle \psi | \hat{V} | \psi\rangle $. 

The reflection and transport probabilities of the junction can be obtained when an incoming fermion hits the interface from the F side. The wave function on the left side of the junction $(x \leq 0)$ is,
\begin{equation}
\psi_L(\textbf{r})=\psi^{+}_{e}(\textbf{r})+r_n\psi^{-}_{e}(\textbf{r})+r_A\psi^{-}_h(r),
\label{Eq.LeftWaveFunction}
\end{equation} 

 The $r_n$ and $r_A$ are  the amplitude of noraml and Andreev reflections, respectively. The wave function on the superconducting side of the junction $(x\geq 0)$ is

\begin{equation}
\psi_R(\textbf{r})=t_e \psi_{e}^{S,+}(\textbf{r})+t_h \psi_{h}^{S,+}
\label{Eq.RightWaveFunction}.
\end{equation}
Here, $t_e$ and $t_h$ are the transmission amplitudes of electron-like and hole-like states on the superconducting side. The $\psi^S_e(\textbf{r})$ and $\psi^S_h(r)$ are the corresponding wave functions of electron-like and hole-like excitations in the superconducting side. We use heavily doped approximation, $\mu_S \rightarrow \infty$, in the S region to satisfy the necessary density of states of induced superconductivity. The boundary condition that matches the wave functions
of two sides of the junction is  necessary to calculate the reflection and
transmission amplitudes\cite{Mondal2010PRL,Yokoyama2009PRL,Yokoyama2010PRB,Linder2010PRB,Linder2010PRL},

\begin{equation}
\psi_L(x=0)=\psi_R(x=0).
\label{Eq.BoundaryCondition}
\end{equation}

The amplitude of Andreev reflection is,
\begin{equation}
r_A=\frac{ie^{i \frac{\theta-\alpha}{2}} \sqrt{\cos\theta \cos\alpha}}{\cos\beta\cos(\frac{\theta-\alpha}{2})+i\sin\beta\cos\frac{\theta+\alpha}{2}}.
\label{Eq.rA}
\end{equation}
Also, the amplitude of normal reflection is,
\begin{equation}
r_e=\frac{e^{i\theta}\left(-\sin\beta\sin(\frac{\theta-\alpha}{2})+i \cos\beta \sin(\frac{\theta+\alpha}{2})\right)}{\cos\beta\cos(\frac{\theta-\alpha}{2})+i\sin\beta\cos\frac{\theta+\alpha}{2}}.
\label{Eq.re}
\end{equation}

The probability of Andreev reflection is $R_A=|r_A|^2$. Also, the other probabilities can be obtained by multiplication of its value to its complex conjugate. These probabilities are useful to calculate the transport properties of the junction.

We define $\Psi(\textbf{r})$ and $\Psi^\dagger(\textbf{r})$ as the annihilation and creation field operators in the Nambu space for Eq.(\ref{Eq.HBdG}). We have $\Psi(\textbf{r})=\left(\Phi(\textbf{r}), \Xi(\textbf{r})\right)^T$, where $\Phi(\textbf{r})=\left(\phi_\uparrow(\textbf{r}), \phi_\downarrow(\textbf{r}) \right)$ is its electron part. The hole part of the field operator is defined as $\Xi(\textbf{r})=\mathcal{T}\Phi(\textbf{r})$ to satisfy the s-wave character of superconductivity.

Using these operators, the $\alpha$-component of spin density wave on the surface of 3D TI is $S_\alpha=\Psi^\dagger \eta_0\sigma_\alpha \Psi$. We rewrite the BdG equation, Eq.(\ref{Eq.HBdG}),  in the real space as:

\begin{equation}
\mathcal{H}_{BdG}=\int d^2r \Psi^\dagger(\textbf{r})H_{BdG} \Psi(\textbf{r}).
\label{Eq.rHBdG}
\end{equation}

We need the commutators of field operators with the Hamiltonian of Eq.(\ref{Eq.rHBdG}) to obtain the dynamics of spin density wave,

\begin{equation}
\begin{array}{l}
\left[\mathcal{H}_{BDG}, \Phi(\textbf{r})\right]=-H_D(\textbf{k})\Phi(\textbf{r})-\sigma_0\Delta_0 \Xi(\textbf{r}),\\
\\
\left[\mathcal{H}_{BdG}, \Phi^\dagger(\textbf{r})\right]=\Phi^\dagger(\textbf{r})H_D(-\textbf{k})-\Xi^\dagger(\textbf{r})\sigma_0\Delta_0^*.
\end{array}
\label{Eq.CommutatorHPsi}
\end{equation}

The commutators of $\Xi(\textbf{r})$ can be calculated in a similar way. The Heisenberg equation of motion can be written for any time-independent operator such as $\hat{A}$,

\begin{equation}
\partial_t\hat{A}=i \left[H , \hat{A} \right].
\label{Eq.Heisenberg}
\end{equation},

This can be used to calculate the dynamics of $S_z$ as below,

\begin{equation}
\begin{array}{rl}
\partial_t S_z =&\left(\partial_t \Psi^\dagger(\textbf{r})\right)\eta_0 \sigma_z \Psi(\textbf{r})+\Psi^\dagger(\textbf{r})\eta_0 \sigma_z \left(\partial_t \Psi(\textbf{r})\right)\\
& \\
 =&\left(\partial_t \Phi^\dagger(\textbf{r})\right)\sigma_z \Phi(\textbf{r})+\Phi^\dagger(\textbf{r})\sigma_z\left(\partial_t \Phi(\textbf{r})\right)\\
 & \\
 + & \left(\partial_t \Xi^\dagger(\textbf{r})\right)\sigma_z \Xi(\textbf{r})+\Xi^\dagger(\textbf{r})\sigma_z\left(\partial_t \Xi(\textbf{r})\right).
 \end{array}
\label{Eq.SD1}
\end{equation}

The commutation relations of Eq.(\ref{Eq.CommutatorHPsi}) can be used to have,
\begin{equation}
\begin{array}{rl}
\partial_t S_z= & i \left[\mathcal{H}_{BdG}, \Phi^\dagger(\textbf{r})\right] \sigma_z \Phi(\mathbf{r}) + \Phi^\dagger(\mathbf{r}) \sigma_z i \left[\mathcal{H}_{BdG}, \Phi(\textbf{r})\right]\\
& \\
+ & i \left[\mathcal{H}_{BdG}, \Xi^\dagger(\textbf{r})\right] \sigma_z \Xi(\mathbf{r}) + \Xi^\dagger(\mathbf{r}) \sigma_z i \left[\mathcal{H}_{BdG}, \Xi(\textbf{r})\right]\\.
\end{array}.
\label{Eq.SD2}
\end{equation}

 The straightforward algebra leads to the dynamics of $S_z$ as:
  
\begin{equation}
\begin{array}{ll}
\partial_t S_z+  \boldsymbol{\nabla}. \textbf{J}^s_z = dT_z.
\end{array}
\label{Eq.SD3}
\end{equation}

Where, $J^s_z=i \hat{e}_z\times\textbf{J} $ is the $z$-component of spin density current and $dT_z=2 \Psi^\dagger\eta_z (\textbf{m}\times \boldsymbol{\sigma}).\hat{e}_z\Psi$ is the density of ATT.

 The spin density wave is independent of time in the steady state approximation, $\partial_t S_z=0$. So, the integration of Eq.(\ref{Eq.SD3}) leads to the $z$-component of ATT as below,

\begin{equation}
T_z= \int dT_z= \int \boldsymbol{\nabla}.\textbf{J}^s_z d^3 r.
\label{Eq.CTT1}
\end{equation}

One can use the divergence theorem  to convert the volume integral into boundary one and rewrite it,

\begin{equation}
T_z=\oint ( \hat{\textbf{e}}_z \times \textbf{J}).d\textbf{l}
\label{Eq.CTT2}.
\end{equation}

 The $d\textbf{l}$ is carried over a closed loop. Since the current is conserved, the Eq.(\ref{Eq.CTT2}) demonstrates that the ATT is related to current bending. Its reduction in one direction leads to an increase in the other. So, the ATT is related to the transverse current that flows parallel to the interface. Due to the absence of $\sigma_z$ in the Eq.(\ref{Eq.Hamiltoina}) and the 2D nature of the junction, the other components of ATT are zero.

\begin{figure}
\includegraphics*[scale=0.31]{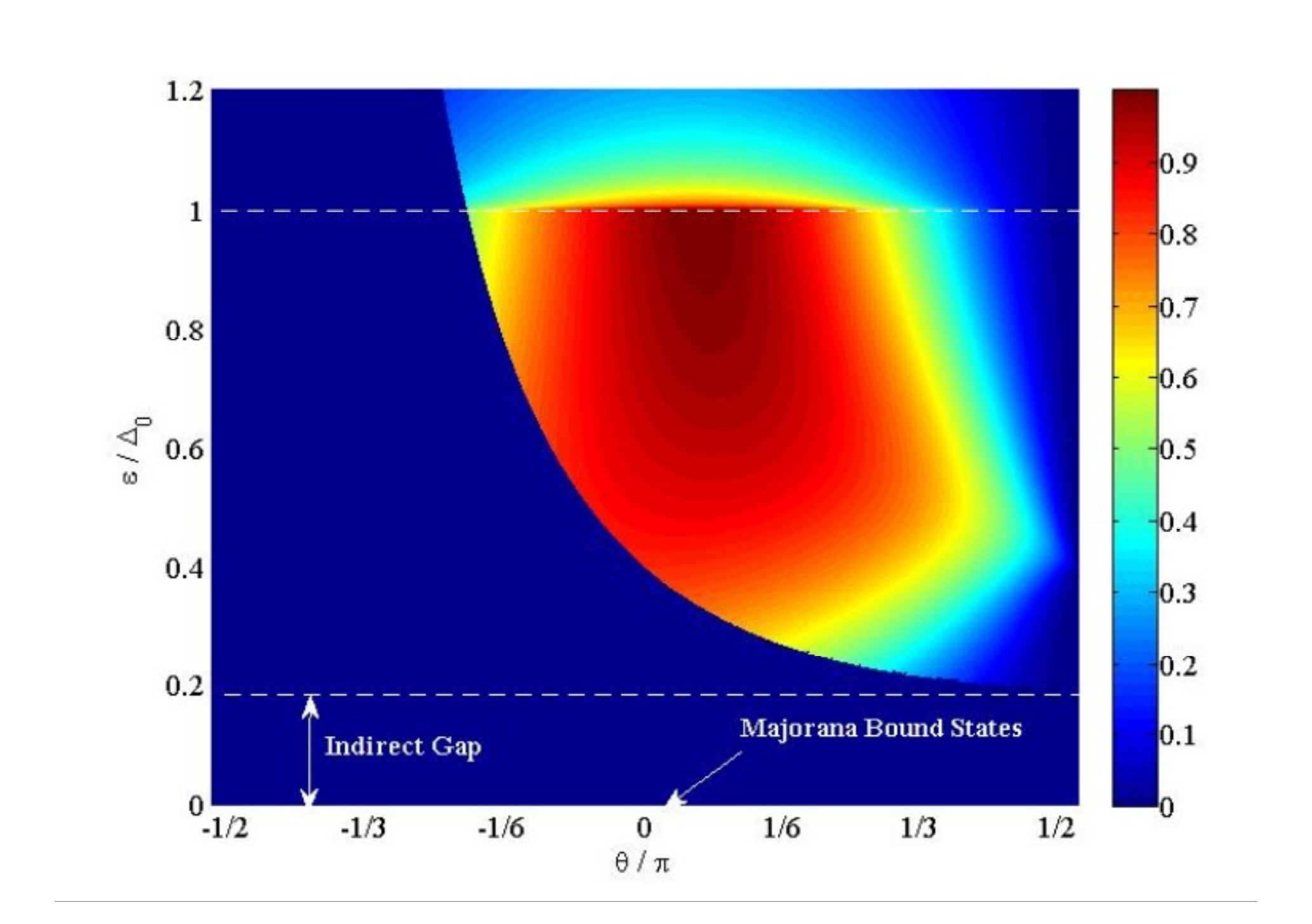}
\caption{(color online) The anisotropic angle-dependent Andreev probability with excitation energy $\epsilon$ normalized by the magnitude of superconducting gap, $\Delta_0$. Here, $\theta$ is the propagation angle of incoming fermion. We set magnetization direction perpendicular to the interface, $m_x=-0.2 \Delta_0$. In the range of $\epsilon \leq m_0$, the Andreev probability is zero. This effect creates an indirect gap that induced by magnetization in the junction.In the above of indirect gap, there is a difference in the probabilities of upward and downward propagating fermions that creates the transverse conductance}.
\label{Fig.Transmission}
\end{figure}

\section{Results and discussion}
\label{Sec.Results}

\subsection{Anisotropic angle-dependent Andreev reflection}
\label{Sec.Re1}
We set the ferromagnetic exchange field in the x-direction, $\textbf{m}=(-m_0,0,0)$, to have negative transverse current in the Andreev dominant regime. 
Also, we set $\mu=0$ and normalize the energy values with respect to $\Delta_0$.
The incoming fermions which belong to the half part of the
electron-like circle with $k_x \geq 0$  hit
the interface from F side of the junction. The hole-like cone located at $\{0,m_0\}$ in the $k$-space. The incoming fermions with $\epsilon \leq \Delta_0$ can not transport to the states at the above of superconducting gap. Since
the radius of each energy circle is $\epsilon$, the minimum energy for incoming fermion is $\epsilon=m_0$ to find a steady states on the hole-like circle. It means an indirect gap induced in the junction in the range of $0 \leq \epsilon \leq m_0$. As illustrated in Fig.(\ref{Fig.Transmission}), all incoming fermions reflect, and $R_A=0$ creates a no-current area in the Andreev reflection probability figure. The electron-hole duality of the BdG wave functions, make them a suitable candidate to be host of the Majorana-bound states. These fermionic states describes particles that can be simultaneously their anti-particles\cite{Fu2010PRL,Yokoyama2009PRL,Linder2010PRL}.
The TI-based superconducting region hosts the Majorana bound states at the $\epsilon=0$. The perfect Andreev reflection, $R_A=1$, and the robust conductance peak at  $G(\epsilon=0)=2G_0$ are their experimental signatures\cite{Salehi2017SciRep}. As shown in Fig.(\ref{Fig.Transmission}) and Fig (\ref{Fig.GL}), the in-plane magnetization sets the Majorana-bound states into the indirect gap, and their signatures disappear. Since magnetization induced by proximity effect in our device, this effect can act as a on/off switch for Majorana-bound states in future technology.

The two energy circles overlaps for values bigger than the indirect gap threshold, $\epsilon   \geq m_0$. The incoming fermion can find a steady-state on the hole-like circle during the Andreev process. So, the propagation direction of reflected hole determines with respect to its location on the hole-like circle. The probability of Andreev reflection becomes sensitive to the propagation angle of incoming fermions, $\theta$. As shown in Fig.(\ref{Fig.Transmission}), incoming fermions with the same energy have different probabilities with respect to its propagation angle. These probabilities grow with respect to energy values, and the perpendicular fermions have a chance more than others. This phenomenon leads an imbalance between upward and downward fermions and creates a transverse current that flows parallel to the interface. This current can be detected via a two extra leads located aside of the junction. The electron-like and hole-like quasi-particles are two types of carriers in our device. The sign of transverse current determines the carrier dominant. We use $m_0 \leq 0$ to have negative value for transverse conductance in hole-like dominant regime and positive value for electron-like dominant regime.
For incoming fermions with $\epsilon \geq \Delta_0$, the states at the above of superconducting gap is accessible. So, the Andreev probabilities encounter a major reduction in their values. This means the transverse conductance approaches to zero for $\epsilon \gg \Delta_0$.

\subsection{Transverse Conductances}
\label{Sec.Re2}

\begin{figure}
\includegraphics*[scale=0.35]{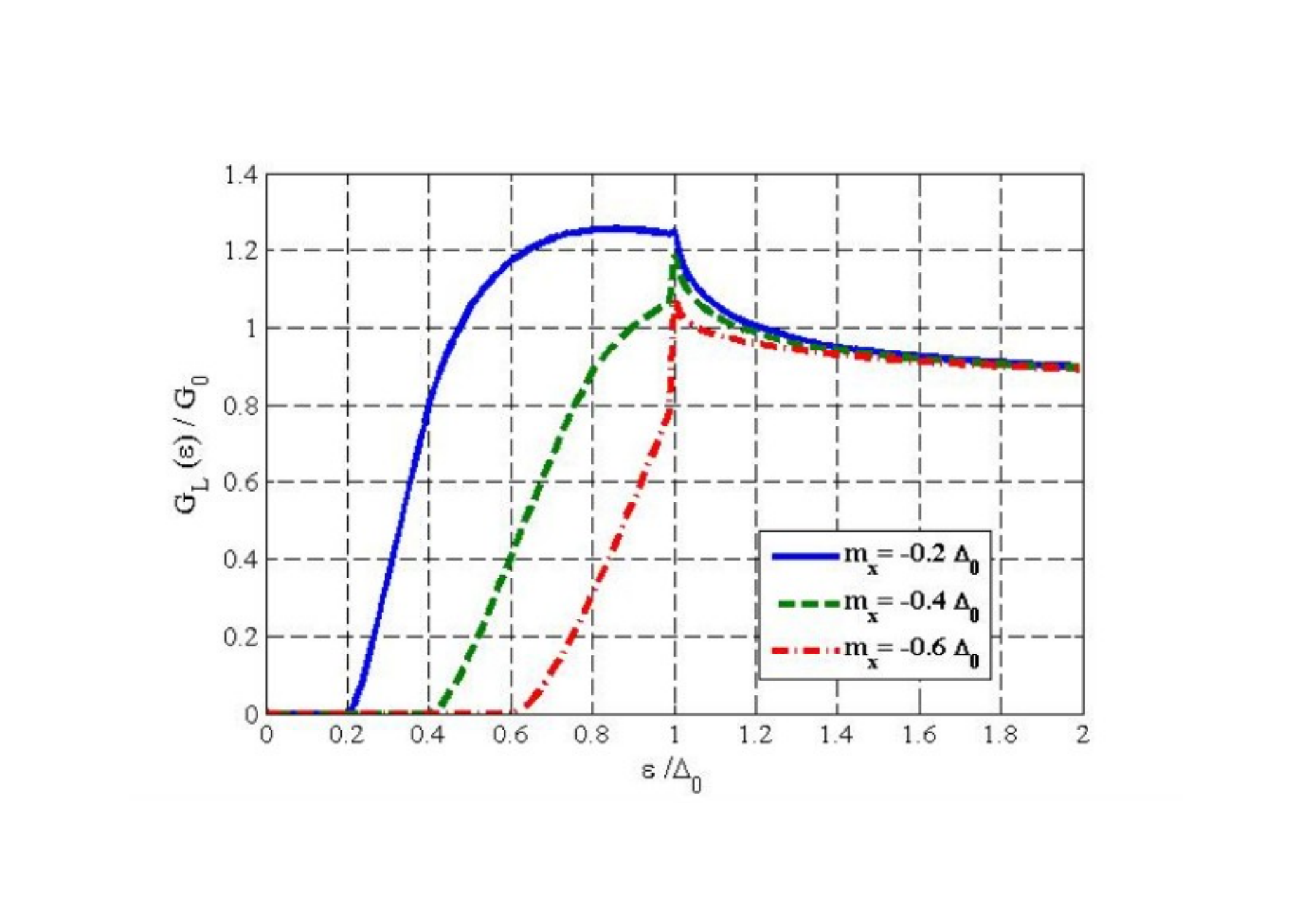}
\caption{(color online) The longitudinal conductances vs. excitation energy. The longitudinal conductance reaches to its maximum in $\epsilon \simeq \Delta_0$, where the Andreev reflection is dominant. The $G_L(\epsilon)$ grows exponentially in the energy range above the indirect gap and approaches to its ballistic value in the high energy regime where the effect of superconducting gap can be ignored. }.
\label{Fig.GL}
\end{figure}
 
 Using transport probabilities of Eq.(\ref{Eq.BoundaryCondition}), one can calculate the transverse and longitudinal conductances \cite{BTK} in x and y directions, respectively. 
 
 \begin{equation}
 G_T(\epsilon)=G_0 \int_{-\frac{\pi}{2}}^{\frac{\pi}{2}} \left(R_N(\theta)\sin\theta-\mathcal{S} R_A(\theta)\sin\alpha\right) d\theta,
 \label{Eq.HallConductance}
 \end{equation}
 
 \begin{equation}
 G_L(\epsilon)=G_0 \int_{-\frac{\pi}{2}}^{\frac{\pi}{2}} \left((1-R_N(\theta))\cos\theta+R_A(\theta) \cos\theta\right)d\theta,
  \label{Eq.LongitudinalConductance}
 \end{equation}
where $G_0=(e^2/h)N(\epsilon)$ is the ballistic conductance and $N(\epsilon)=(|\epsilon+\mu|W)/\hbar v_F$ is the density of states. Also, $W$ is the width of the junction.

In contrast to Graphene\cite{Beenakker2006PRL,Sengupta2006PRL,Linder2007PRL} and other relativistic materials such as TIs \cite{Yokoyama2010PRB,Linder2010PRB} or Weyl semimetals where the zero-bias conductance peak and Majorana-bound states are responsible for sub gap conductances, the no-current domain originates from the indirect gap. As shown in Fig.(\ref{Fig.GL}), the longitudinal conductance arises exponentially at $\epsilon=m_0$ and tends to its maximum value at the edge of superconducting gap, where the Andreev probability is dominant. In high energy regime, the effect of Dirac cone displacement  and superconducting gap can be ignored and the $G_L$ approaches to its ballistic value. As shown in part (a) of Fig (\ref{Fig.1}), the longitudinal conductance passes through the battery and can be measured by an ammeter.

\begin{figure}
\includegraphics*[scale=0.43]{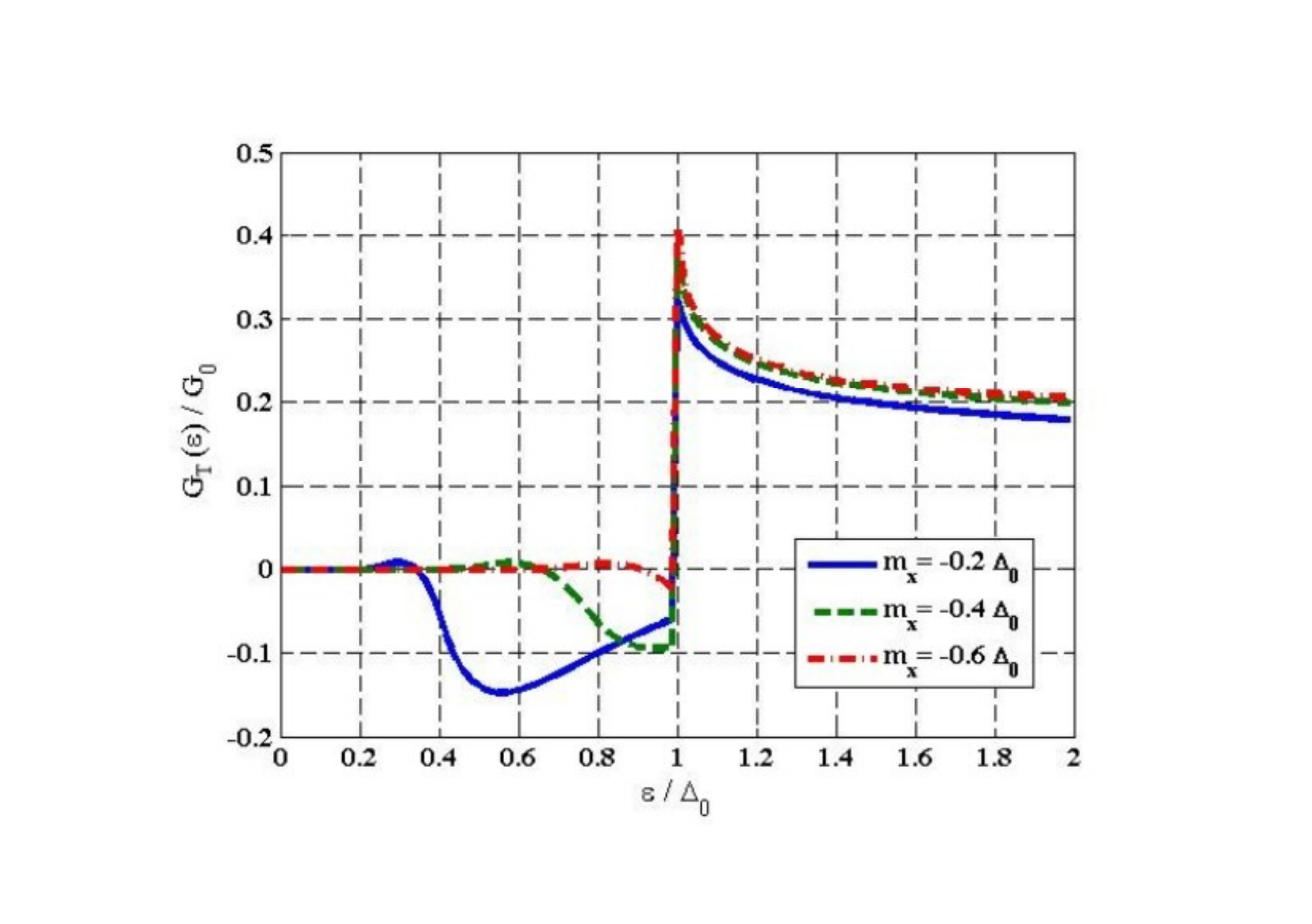}
\caption{(color online) The transverse conductance vs. energy excitation for different values of magnetization direction. This type of transverse conductance originates from the scattering at the interface. }
\label{Fig.GH}
\end{figure}

The transverse conductance is demonstrated in Fig.(\ref{Fig.GH}). Based on the Eq.(\ref{Eq.HallConductance}), those propagation directions with perfect normal reflection, $\{R_N=1, R_A=0\}$, have no effect on the $G_T$. Because of $s$-wave character of superconductivity, the states above the superconducting gap in the S side of the junction have no effect on the $G_T$. Also, each integral can be considered into two parts, the downward section with $-\pi/2 \leq \theta \leq 0$ and upward one with $ 0 \leq \theta \leq \pi/2$. The normal reflection positively contributes to the $G_T$ from the upward part, whereas its contribution is negative from the downward part. This term is a little complicated for the reflected hole because it has two types of retro and specular reflections. We use the $\mathcal{S}$ term to take this into account. In the absence of chemical potential, $\mu=0$, we have $\mathcal{S}=1$. These parts compete to determines the sign of $G_T$. The $G_T$ changes from zero at $\epsilon=m_0$, where the incoming fermion can find an spot on the hole-like cone during the Andreev process. Since Andreev reflection is dominant in the range of $m_0 \leq \epsilon \leq \Delta_0$, the transverse conductance is negative. The $G_T$ reaches its minimum around $\epsilon=2 m_0$ where the normal reflection starts to overcome the Andreev one. The damping of Andreev reflection at the edge of superconducting gap leads to the positive value for $G_T$. This effect can be used in a real experiment to determine the value of superconducting gap. Since the physics of $G_T$ originates from the relative location of electron-like and hole-like cones, it tends to zero at the high energy regime , $\epsilon \gg \Delta_0$, where this effect  can be ignored. Also, the sign of $G_T$ depends on the Dirac cone's location. The ferromagnetism induced into the surface of 3D TI by means of the  proximity effect. This occurs by a ferromagnetic lead such as $MnO_2$ or $EuO$. So, the rotation of the ferromagnetic lead rotates the direction of magnetization and  changes the sign of $G_T$ and can be detected in the Hall voltage experiment.

\subsection{Andreev Transfer Torque}
\label{Sec.Re3}

To obtain the $z$-component of ATT, we consider a square with side length of A on the junction. According to the Eq.(\ref{Eq.CTT2}), the incoming fermion enters to the square with propagation direction of $\theta$. During the Andreev process it reflects as a hole quasi-particle and leaves the square with the probability of $R_A$ and different propagation direction of $\alpha$.  We want to obtain the net current which passes through these mutually parallel lines. From Eq.(\ref{Eq.CTT2}), we have,

\begin{equation}
T_z=\oint (J_y dx-J_x dy)=A ( \delta G_y-\delta G_x) 
\label{Eq.CTT3}
\end{equation}

where $\delta G_y \sim G_T$ is difference between conductances flows in the presence and absence of in-plane magnetization parallel to the junction, respectively. The probability conservation dictates the incoming current density into the square to be equal with the outgoing one. So, we have  $\delta G_x=-\delta G_y$ and the final result is,

\begin{equation}
T_z \sim 2 G_T A.
\label{Eq.CTTFinal}
\end{equation}
Here, $T_z$ directly related to the transverse conductance. The absence of $\sigma_z$ term in the spin-orbit coupling term of Eq.(\ref{Eq.Hamiltoina}) and two-dimensional nature of TI-based junction lead the other components of ATT, $\{T_x,T_y\}$, to zero. The ATT imposes on the junction positively and negatively corresponds to positive and negative transverse conductances, respectively. In the energy range that transverse conductance is maximum, the ATT reaches to its maximum, too. This originates in the bending of propagation direction. These effects in the energy range of $m_0 \leq \epsilon \leq  \Delta_0$ are important and experimentally detectable. Moreover, the ATT approaches zero in the high energy limit where the Dirac cone's displacement can be ignored.

\section{Conclusion}
\label{Sec.Concl}
Lets discuss about the output prediction of our model. The typical value of induced ferromagnetic field by means of proximity effect on the surface of 3D TIs is $5 \sim 50 meV$. This occurs by a ferromagnetic electrode such as $EuO$ or $MnO_2$ that deposits on the surface of $Bi_2 Se_3$ \cite{Haugen2008PRB}. Also, the induced superconducting gap has the same order of ferromagnetic field. The high-quality topological insulators that can be fabricated now, have sufficiently large coherence length up to $\sim 370 nm$. It means the boundary details and localization effect are negligible, and the ballistic limit is a well approximation to explore the junction \cite{Yokoyama2010PRB}. We normalized our results with respect to the superconducting gap. It means the approximated value of the indirect gap would be $2.5 \sim 25 meV$. Also, the usual width of the junction in a real situation is of the order of the coherence length. So, the approximated value of the transverse conductance would be $\sim 10^{-4} Simens$, which can be measurable with available technology. Finally, the magnitude of ATT is $\sim 10^{-4} Simens (\mu m)^2$.

In this paper, we consider the effect of angle-dependent Andreev reflection of an F/S junction on the surface of 3D TIs. We show whenever in-plane magnetization has a component perpendicular to the interface, the electron-like and hole-like cones separate in the $k$-space. This induces an indirect gap in the junction. Also it creates a no-current area in longitudinal conductance. Moreover, the Andreev reflection is angle-dependent that leads to a transverse conductance. Further, we show the sign of transverse conductance is related to the magnetization direction. Based on this, we design an experimental set up to reveal it. At las but not least, we illustrate that the transverse conductance imposes a torque on the junction that is important in the low-energy limit near the Dirac point. The $z$-component of this torque, ATT, is non-zero, and its sign and magnitude are related to the sign and magnitude of the transverse conductance. Since the STT is very important for data storage technology, we sure its extension to ATT can be very useful in designing and fabricating new devices for further applications.

\end{document}